\documentclass[useAMS,usenatbib]{mn2e}
\usepackage{graphicx,graphics,amsmath,amssymb,mathrsfs}

\newcommand{\sd}{\Sigma} 
\newcommand{\pom}{\dot{\varpi}}

\newcommand{\ddr}{\frac{d}{dR}}  

\newcommand{\DDR}{\frac{\partial}{\partial{R}}}

\newcommand{\der}[2]{\frac{d{#1}}{d{#2}}}
\newcommand{\DER}[2]{\frac{\partial{#1}}{\partial{#2}}}
\newcommand{\rs}{R_{\star}} 
\newcommand{\drr}{\frac{d^2}{dR^2}}
\begin{document}
\title {Slow pressure modes in thin accretion discs}

\author[Saini, Gulati \& Sridhar]{Tarun Deep Saini$^{1,3}$, Mamta Gulati$^{1,2,4}$, S. Sridhar$^{2,5}$  \\
  $^{1}$ Indian Institute of Science, Bangalore, Karnataka, India, 560 012 \\
  $^{2}$ Raman Research Institute, Sadashivanagar, Bangalore, Karnataka, India, 560 080 \\
  $^{3}$ tarun@physics.iisc.ernet.in \\
  $^{4}$ mgulati@rri.res.in  \\
  $^{5}$ ssridhar@rri.res.in 
} 
\maketitle

\begin{abstract}
Thin accretion discs around massive compact objects can support slow
pressure modes of oscillations in the linear regime that have
azimuthal wavenumber $m=1$. We consider finite, flat discs composed of
barotropic fluid for various surface density profiles and
demonstrate--through WKB analysis and numerical solution of the
eigenvalue problem--that these modes are stable and have spatial
scales comparable to the size of the disc. We show that the eigenvalue
equation can be mapped to a Schr\"odinger-like equation. Analysis of
this equation shows that all eigenmodes have discrete spectra. We find
that all the models we have considered support negative frequency
eigenmodes; however, the positive eigenfrequency modes are only
present in power law discs, albeit for physically uninteresting values
of the power law index $\beta$ and barotropic index $\gamma$.
\end{abstract}


\section{Introduction}
Low-mass discs orbiting massive compact bodies are a feature of many
astronomical systems. When the dynamics of a disc is dominated by the
Newtonian gravitational force of the central body, the disc may be
considered nearly Keplerian.  In a purely Keplerian potential
eccentric orbits do not precess because the orbital frequency is equal
to the epicyclic frequency. In a nearly Keplerian disc there is a
small difference between the orbital and epicyclic frequencies. This
could be due to the self-gravity of the disc, thermal pressure in a
gas disc, and random motions in a collisionless disc. This difference
in frequencies manifests as a precession of eccentric orbits at rates
that are small compared to the orbital and epicyclic frequencies. Then
the disc may be able to support large-scale, slow, lopsided modes
\citep{kat83, sst99, lg99, st99}.  In the linear regime, these modes
have azimuthal frequency, $m=1$, whose first systematic investigation
is due to \citet{tre01}. He studied slow modes in various types of
discs (fluid, collisionless and softened gravity), with the focus
largely on the effect of the self-gravity of the disc. In particular,
a WKB analysis was used to show that the fluid disc can support
large-scale slow modes when the Mach number, ${\mathscr M}$, is much
larger than the Toomre $Q$ parameter (both parameters are defined in
\S~2). The assumption behind this analysis is that the self-gravity of
the disc dominates fluid pressure.  However, such is not the case for
thin accretion discs around white dwarfs and neutron stars. Indeed,
for a disc around a white dwarf \citep{fkr02}, we can estimate
${\mathscr M}\sim 50$ and $Q\sim 10^{10}$; hence the analysis of
\citep{tre01} is not directly applicable. The goal of this paper is to
study large-scale $m=1$ slow modes in thin accretion discs, where
$Q\gg {\mathscr M}\gg 1$.

In \S~2 we use the WKB approximation to establish that pressure, in
the absence of self-gravity, can enable slow, $m=1$ modes in thin
accretion discs. The linear eigenvalue problem for slow pressure modes
(or ``p-modes'') is formulated in \S~3 for a flat barotropic disc,
which is axisymmetric in its unperturbed state. When appropriate
boundary conditions are chosen, the eigenvalue equation reduces to a
Sturm-Liouville problem.  Since the differential operator is
self-adjoint, the eigenfrequencies are real: therefore all p-modes are
stable, and the eigenfunctions form a complete set of orthogonal
functions. We also map the eigenvalue equation into a
Schr\"odinger-like equation, which is useful in the interpretation of
our numerical results. In \S~4 we present numerical results for a
variety of discs, namely (i) an approximation to the Shakura-Sunyaev
thin disc, (ii) the Kuzmin disc which is more centrally concentrated,
(iii) power-law discs. Of particular interest is the nature of the
eigenfrequency; whether it is positive or negative.  This has bearing
on the excitation of these modes, because they are stable
and will not grow spontaneously through a non-viscous
instability. Comparison with earlier work, summary and conclusions
follow in \S~5.

\section{Slow pressure modes}

Consider a flat thin disc of fluid orbiting a central mass $M$. The
fluid is assumed to be barotropic and the disc is described by a surface 
density profile $\sd$.  In our analysis we ignore viscous
forces, assuming that they adjust to maintain a quasi-stationary flow
with a small radial velocity, and have little effect on the perturbed
flow. Thus, we start with the continuity and Euler equations in
cylindrical polar coordinates
\begin{eqnarray}
&&\DER{\Sigma}{t}+ \frac{1}{R}\DER{}{R}(R\Sigma v_R) + \frac{1}{R}\DER{}{\phi}(\Sigma v_{\phi}) = 0\,,
\nonumber\\
&&\DER{v_R}{t}+v_R\DER{v_R}{R}+\frac{v_{\phi}}{R}\DER{v_R}{\phi}-\frac{{v_{\phi}}^2}{R}= 
- \frac{GM}{R^2} -\DER{}{R} 
(\Phi+h)\,,\nonumber\\
&&\DER{v_{\phi}}{t}+v_R\DER{v_{\phi}}{R}+\frac{v_{\phi}}{R}\DER{v_{\phi}}{\phi}+\frac{{v_R}{v_{\phi}}}{R}
= -\frac{1}{R} \DER{}{\phi}(\Phi+h)\,,
\label{eq:euler}
\end{eqnarray}
where $v_R$ and $v_{\phi}$ are the radial and azimuthal components of
the fluid velocity, $h$ is the enthalpy per unit mass, and $\Phi$ is the gravitational potential
due to the disc. For a barotropic fluid with an equation of state given by
$p=D{\Sigma}^{\gamma}$ (where $D>0$ is a constant), the isentropic sound
speed and enthalpy are given by
\begin{eqnarray}
&&{c_s}^2 = \gamma\,D\,\Sigma^{\gamma-1}\,,\\
\label{eq:enthalpy} 
&&h=\frac{\gamma{D}}{\gamma-1}{\Sigma}^{\gamma-1} =\frac{c_s^2}{\gamma-1}\,,
\end{eqnarray}

\subsection{Precession rate in the unperturbed disc}
We assume that the radial velocity of the unperturbed flow is much
smaller than the azimuthal velocity and set it identically equal to
zero; this assumption is justified below at the end of \S~2.2.  The
unperturbed disc is assumed to be axisymmetric, therefore all $\phi$
derivatives are set to zero. Gas flows along circular orbits, with
centrifugal balance maintained largely by the gravitational attraction
of the central mass (with small but non trivial contributions from gas
pressure and disc self-gravity). The azimuthal and radial frequencies,
$\Omega >0$ and $\kappa >0$ respectively, associated with nearly
circular orbits are given by
\begin{eqnarray} 
\Omega^2 &=& \frac{GM}{R^3} + \frac{1}{R}\ddr(\Phi_0 + h_0)\,,\nonumber\\
\kappa^2 &=& \frac{GM}{R^3} + \frac{d^2}{dR^2}(\Phi_0 + h_0) + \frac{3}{R}\ddr(\Phi_0 + h_0)\,,
\label{eq:freqs}
\end{eqnarray}
where the subscript `$0$' indicates unperturbed quantities.  The Mach
number of the flow ${\mathscr M}(R)= R\Omega(R)/c_s(R) \gg 1$.  The
dominant contribution to both $\Omega^2$ and $\kappa^2$ is due to the
central mass, with small corrections coming from the disc self-gravity
$(\Phi_0)$ and enthalpy $(h_0)$. Let us define the small parameter
$\epsilon\ll 1$ as the larger of $(\Sigma_0 R^2/M)$ and
$(h_0R/GM)$. The apsides of the nearly circular orbit of a fluid
element precesses at a rate given by,
\begin{eqnarray}
\pom &=& \Omega - \kappa\nonumber\\
&=& -\frac{1}{2\,\Omega} \left( \frac{d^2}{dR^2} + \frac{2}{R}\der{}{R}\right)
\left(\Phi_0 + h_0\right) + O(\epsilon^2)\,.
\label{eq:pomegadot}
\end{eqnarray}
Note that, in contrast to \citet{tre01}, we have retained the
contribution from gas pressure (i.e. enthalpy). In fact, in thin
accretion discs around compact stars, disc self-gravity is negligible
and the contribution to $\pom$ is almost entirely from gas
pressure. The goal of this section is to establish that these discs
have large-scale slow modes driven only by pressure. In the WKB
analysis of linear modes given below we follow the presentation due to
\citet{tre01}.

\subsection{The WKB approximation} 
We consider linear perturbations (of the velocity, surface density
etc) of the form
\begin{equation}
A(R)\exp{\left[i\left(\int^R k(R) dR + m\phi - \omega t\right)\right]}\,,
\end{equation} 
where $k(R)$ and $m$ are the radial and azimuthal wavenumbers,
respectively, and $\omega$ is the angular frequency of the mode. In
the tight-winding limit where $\vert k(R)R\vert\gg 1$, a dispersion
relation between $\omega$ and $k(R)$ can be derived \citep{saf60,
  bt08}:
\begin{equation} 
\left(\omega - m\Omega\right)^2 = \kappa^2 - 2\pi G\Sigma_0\vert k\vert +
c_s^2k^2\,.
\label{eq:drfull}
\end{equation}
The disc is stable to axisymmetric $(m=0)$ perturbations if and only if
\begin{equation}
Q\equiv \frac{c_s\kappa}{\pi G\Sigma_0} > 1\,.
\label{eq:q}
\end{equation}
This is readily satisfied in thin accretion discs around compact
stars.  \citet{tre01} showed that the dispersion relation for $m=1$
modes can be written as
\begin{equation}
\omega = \pom + \frac{\pi G\Sigma_0\vert k\vert}{\Omega} - 
\frac{c_s^2k^2}{2\Omega} + \frac{1}{\Omega}O(\pom^2, \omega^2)\,.
\label{eq:dr}
\end{equation}

In \citet{tre01} it is argued that, when the pressure is negligible
compared to disc self-gravity (i.e. $c_s\approx 0$), the WKB
dispersion relation of equation~(\ref{eq:dr}) admits large-scale
`$\vert k(R)R\vert \sim 1$' modes with frequencies $\omega\sim\pom\sim
(\Sigma_0 R^2/M)\Omega$. It may be verified that the condition of
negligible pressure implies that the Mach number ${\mathscr M}\gg Q$.
However, as we have argued in the introduction, this inequality is
violated for thin accretion discs around compact stars where the
opposite is true, i.e. $Q\gg {\mathscr M}\gg 1$. Hence we need to
consider a situation that is complementary to the analysis of
\citet{tre01}. Disc self-gravity being negligible in accretion discs,
the precession rate is determined entirely by the gas pressure. Then
equation~(\ref{eq:pomegadot}) can be written as
\begin{equation}
\pom = -\frac{1}{2\,\Omega} \left[\frac{d^2h_0}{dR^2} + 
\frac{2}{R}\der{h_0}{R}\right] + O(1/{\mathscr M}^4) \sim \frac{\Omega}{{\mathscr M}^2}\,,
\label{eq:pomh}
\end{equation}
(see \citet{kat83}) and we can approximate equation~(\ref{eq:dr}) as
\begin{equation}
\omega = \pom  - \frac{c_s^2k^2}{2\Omega} + \frac{1}{\Omega}O(\pom^2, \omega^2)\,.
\label{eq:drslow}
\end{equation}
For a disc with a non zero inner radius, eigenmodes must
satisfy the Bohr-Sommerfeld quantization condition, given by
\begin{equation}
\oint k(R)\,dR = \left(n+\frac{3}{4}\right)2\pi \,,\qquad n=0,1,2,\ldots
\label{eq:quant}
\end{equation}
Thus, there exists a {\it prima facie} case for modes with frequencies 
$\omega\sim\pom$, with radial wavenumbers,
\begin{equation}
k(R)\sim \left\vert\frac{\Omega\pom}{c_s^2}\right\vert^{1/2} \sim \frac{1}{R}\,,
\label{eq:k}
\end{equation}
comparable to the radial scale of the disc.  However, this tentative
conclusion is based on a WKB analysis which may not be valid for modes
with $k(R)R\sim 1$. This is the motivation for our studies of the
eigenvalue problem for slow modes given below. Henceforth we ignore
disc self-gravity altogether and consider only the effect of gas
pressure.

We now justify the assumption made at the beginning of \S~2.1, that
the radial velocity in the unperturbed disc is small, and may be
ignored when studying slow modes.  The time scale of radial spreading
of the disc is, $t_{vis}\sim R^2/\nu \sim
\mathscr{M}^{2}/(\alpha\Omega)$, where $\alpha \ll 1$ is the
Shakura--Sunyaev viscosity parameter. The frequency of a slow mode is
$\omega\sim \pom \sim \Omega\mathscr{M}^{-2}$. Therefore $\omega
t_{vis}\sim \alpha^{-1}\gg 1$ implies that the radial spreading occurs
over many slow mode periods.

\section{Formulation of the eigenvalue problem}
\subsection{Eigenvalue equation}
The linearized Euler, continuity and enthalpy equations that govern
the perturbed flow are
\begin{eqnarray}
&&\der{v_{R1}}{t} - 2\Omega(R){v_{\phi 1}}=-\DER{h_1}{R}\,,
\label{eq:pert1}\\
&&\der{v_{\phi 1}}{t} -2B(R)v_{R1} = -\frac{1}{R}\DER{h_1}{\phi}\,, \label{eq:pert2}\\
&&\der{\sd_1}{t} + \frac{\sd_0}{R}\DER{v_{\phi 1}}{\phi} + \frac{1}{R}\DDR(R\sd_{0}v_{R1}) = 0,\label{eq:pert:cont}\\
&&h_1 = c^2_{s0}\frac{\sd_1}{\sd_0}, \label{eq:pert:eth}
\end{eqnarray}
where the subscript `0' stands for the unperturbed quantities and `1'
for the first order perturbed quantities. The Oort's parameter $B(R)$
is related to the epicyclic frequency through $\kappa^2(R) =
-4\Omega(R) B(R)$, and $d/dt = (\partial/\partial t +
\Omega\partial/\partial\phi)$ is the convective derivative with
respect to the unperturbed flow.

We consider non-axisymmetric perturbations with azimuthal wave number
$m=1$, of the form $T_1=T_{a}(R) \exp\left[ i(\phi-\omega t)\right]$,
where $T_1$ stands for any perturbed quantity. Substituting this form in
equations~(\ref{eq:pert1}), (\ref{eq:pert2}), (\ref{eq:pert:cont}) and
(\ref{eq:pert:eth}) yields
\begin{eqnarray}
&&i(\Omega-\omega){v_{Ra}}-2\Omega{v_{\phi a}}+\der{h_a}{R}=0\,\label{eq:pert:11}\,,\\
&&i(\Omega-\omega){v_{\phi a}}-2B{v_{R a}}+\frac{i h_a}{R} =0\,\label{eq:pert:21}\,,\\
&& i (\Omega - \omega) \sd_a + \frac{i\sd_0}{R} v_{\phi a} +\frac{1}{R} \ddr(R\sd_0 v_{Ra})=0\,,\label{eq:pert:cont1}\\
&& h_a  = c^2_{s0}\frac{\sd_a}{\sd_0}\,.\label{eq:pert:eth1}
\end{eqnarray}
Solving equations (\ref{eq:pert:11}) and (\ref{eq:pert:21}) for the velocity amplitudes we obtain
\begin{eqnarray}
v_{R a} &=& -\frac{i}{\Delta}\left[(\Omega-\omega)\der{h_a}{R}+\frac{2\Omega}{R}h_a\right]\,,\label{eq:pert:12}\\
v_{\phi a} &=& \frac{1}{\Delta}\left[-2B\der{h_a}{R}+\frac{\Omega-\omega}{R}h_a\right]\,,\label{eq:pert:22}
\end{eqnarray}
where $ \Delta = \kappa^2-(\Omega-\omega)^2$. These equations, along with equation (\ref{eq:pert:eth1}), when substituted in (\ref{eq:pert:cont1}) yields
\begin{eqnarray}
&&\Bigg[ \drr + \left\lbrace \ddr \ln\left(\frac{R\sd_0}{\Delta}\right)\right \rbrace \ddr +\nonumber\\
&& 
\frac{2\Omega}{R(\Omega - \omega)}\left\lbrace \ddr \ln\left(\frac{\sd_0 \Omega}{\Delta}\right)\right \rbrace - 
\frac{1}{R^2}\Bigg]h_a = \frac{h_a\Delta}{c^{2}_{s0}}\,.\label{eq:gold}
\end{eqnarray}
which is the eigenvalue problem for {\it undriven modes}, with eigenvalue $\omega$ and 
eigenfunction $h_a$. This equation is a special case of equation (13) of \citet{gold79, tlai09},
where $m = 1$, and the external and self gravity perturbations are set
equal to zero. It can be noted that this equation becomes singular at
$\Omega = \omega$ and $\Delta = 0$. The former corresponds to the
corotation resonance and the latter corresponds to the 
Lindblad Resonances (LR). Below we discuss the validity of equation
(\ref{eq:gold}) at  LR; a similar analysis holds for the singularity at
corotation radius but we do not discuss it in this paper since, as is
argued later, for the slow modes the corotation radius has to lie outside
the disc.

The system of equations (\ref{eq:pert:11})---(\ref{eq:pert:eth1})
describe an undriven, autonomous system.  At the Lindblad resonance
the algebraic equations (\ref{eq:pert:11}) and (\ref{eq:pert:21}),
become indeterminate if no conditions are imposed on the enthalpy
perturbations $h_a$. It is easily seen that these equations become
consistent if
\begin{equation}
 -\frac{i(\Omega - \omega)}{2B} = \frac{2i\Omega}{(\Omega - \omega)} = \frac{dh_a/dR}{ih_a/R}\,.
\end{equation}
The first equality follows from  $\Delta = 0$. Rearranging the second equality yields
\begin{equation}
\left[ \ddr(R^2 h_a) - \frac{\omega}{\Omega}\left(R^2 \der{h_a}{R}\right)\right]\Bigg|_{\rm LR} = 0.\label{eq:ILR:1}
\end{equation}
This condition must be satisfied at the Lindblad resonances for all 
undriven modes. However, equation (\ref{eq:ILR:1}) may not be satisfied if the
disc is driven by external forcing and may lead to curious dynamics
around the LR and transport of angular momentum away from the LR due
to external torquing \citep{gold79}. In this work we
confine our investigations to free modes of an undriven disc,
therefore, as the above discussion shows, nothing special happens
at the LR.

{\bf Slow Mode Approximation}: We now make the ansatz that the
perturbed flow supports frequencies that are small in comparison to
the circular frequency, i.e., $\vert\omega\vert\ll\Omega$. Therefore, 
when $\omega\neq 0$, the disc must be finite, with outer radius such 
that the orbital frequency at the outer edge is much greater than 
$\vert\omega\vert$. Applying the slow mode approximation to equation 
(\ref{eq:gold}) we obtain 
\begin{equation} 
\frac{c^2_{s0}R^{3/2}}{\sd_0 \Omega}\ddr\left(\frac{\sd_0 \Omega}{ R^{3/2} 
\,\Delta}\ddr (R^2 h_a)\right) = R^2 h_a\,.\label{eq:apgold}
\end{equation}
Similar to equation (\ref{eq:gold}), equation (\ref{eq:apgold}) too is
singular at the Lindblad resonances, however the singularity at the
corotation radius has gone away since this equation has been derived
under the slow mode condition, $\vert\omega\vert\ll\Omega$. The
condition $\Delta=0$ implies that at some radius, either
$\omega=\Omega -\kappa$ or $\omega=\Omega +\kappa$.  Since $\kappa
\simeq \Omega + O(\epsilon^2)$, we see that the second equality cannot
be satisfied under the slow mode approximation. It is straightforward
to see that the radius where this would be satisfied would be larger
than the corotation radius due to the fact that the Keplerian circular
frequency falls off monotonically with radius. Therefore, there are no
outer Lindblad resonance singularities for slow modes. However, the
Inner Lindblad Radius (ILR), where $\omega=\pom(R)$, could very well
lie inside the disc. Due to the fact that the disc surface density is
completely arbitrary, there could in general be more than one ILRs. To
make the problem well posed under the slow mode approximation, at the
ILRs, the condition (\ref{eq:ILR:1}) reduces to
\begin{equation}
\ddr{(R^2 h_a)} \bigg |_{\rm ILR} = 0\,.\label{eq:ILR}
\end{equation}
We shall see later that the numerical solutions of equation
(\ref{eq:eigen}) satisfy this condition and the velocity amplitude at the ILR 
remains finite; thus the linear approximation remains valid and nothing special 
happens at the ILR.

From equation (\ref{eq:pomh}), $\pom /\Omega \simeq O({\mathscr
M}^{-2})$.  This allows us to approximate $\kappa^2 \simeq \Omega^2$,
leading to $B\simeq -\Omega/4$. Using these in equation (\ref{eq:pert:22}) we obtain
\begin{eqnarray}
\label{eq:vphi}v_{\phi a} &=& \frac{\Omega}{2\Delta}\left[\der{h_a}{R}+\frac{2h_a}{R}\right]\,,\nonumber\\ 
           &=&  \frac{\Omega}{2\Delta} \frac{1}{R^{2}} \der{}{R} \left (R^{2}h_a \right)\,.
\end{eqnarray}
Differentiating equation (\ref{eq:apgold}) and using equation (\ref{eq:vphi}) we obtain,
\begin{equation}
\label{eq:eigen} \ddr \left[\left(\frac{c^2_{s0} R^{3/2}}{\sd_0 \Omega}\right) \der{\Theta}{R}\right] + \frac{2R^{3/2}}{\sd_0} (\pom-\omega)\Theta = 0\,,
\end{equation} 
where we have used the variable $\Theta=R^{1/2}\sd_0\,v_{\phi a}$  
and $\Delta \simeq 2\Omega(\omega-\dot{\varpi})$, which is valid under the
slow mode approximation.

\subsection{Slow modes as a Sturm-Liouville problem}
Before we proceed to specific examples, we have to choose the boundary
conditions that we impose to solve equation~(\ref{eq:eigen}). We first
cast the equation in a dimensionless form by choosing a radius $\rs$,
at which we evaluate various quantities, $\sd_{\star}, c_{s\star},
\pom_{\star}$ and $\Omega_{\star}$. We introduce the parameter
$x=R/\rs$, and similarly for a quantity $H$, we use $H'=H/H_{\star}$,
leading to the Sturm-Liouville form of the eigen equation
\begin{equation}
\label{eq:sl}\der{}{x}\left({\rm P}(x)\der{\Theta}{x}\right)  +\left({\rm Q}(x) + \lambda {\rm W}(x) \right)\Theta=0\,,
\end{equation}
where 
\begin{align}
{\rm P}(x) & =\frac{{c'_{so}}^2{x^{3/2}}}{\Sigma'_0{\Omega'}}, \quad {\rm W}(x) = \frac{2x^{3/2}}{\Sigma'_0}\,,\nonumber\\
{\rm Q}(x) & = \frac{2x^{3/2}\dot{\varpi'}}{\Sigma'_0}, \quad \Theta =x^{1/2}{v_{\phi_a}}{\Sigma'_0}\,.
\nonumber
\label{eq:pqw}
\end{align}
\noindent
In equation~(\ref{eq:sl}), $\lambda = -\omega
\mathscr{M_{\star}}^2/\Omega_{\star} $ is defined with a negative sign
to make an explicit correspondence with the Schr\"odinger's equation,
to be introduced in \S~3.3. Henceforth we reserve the term
``eigenvalue'' for $\lambda$, and use either ``frequency'' or
``eigenfrequency'' for $\omega$.

We consider discs with an inner edge at $R_{\rm inner}$, and an outer
edge at $R_{\rm outer}$, and we choose $\rs=R_{\rm inner}$. We now
argue that the parameters of the disc and the central mass for
astrophysically interesting discs are such that the slow mode
condition can be easily satisfied everywhere inside the disc. In a Keplerian
disc the slow
mode condition $|\omega|\ll\Omega$ is satisfied everywhere in the
disc if it is satisfied at the disc outer radius, this leads to
\begin{equation}
\left(\frac{R_{\rm outer}}{R_{\star}}\right)^{3/2}\ll\frac{{\mathscr M_{\star}}^2}{|\lambda|}\,,
\end{equation}
where we have used $\lambda =
-\omega\mathscr{M_{\star}}^2/\Omega_{\star} $, and $\Omega(R_{\rm
  outer}) = \Omega_{\star}(R_{\star}/ R_{\rm outer})^{3/2}$.  Typical
expected values for $\mathscr{M_{\star}}$ are in the range
$10^{4}$--$10^{6}$ \citep{fkr02}. Most examples we consider have
surface densities that decline by $R_{\rm outer}/R_{\star}\simeq
30$--$50$, therefore we see that for an eigenmode to be slow through
out the disc, $|\lambda|$ has to be much smaller than $\sim
10^5$--$10^9$. We shall see in the examples that this condition is
comfortably satisfied.

We integrate the eigen equation in the range $1<x<x_{\rm outer}$. We
assume that the perturbations obey the boundary conditions,
\begin{equation}
\Theta(1)=\Theta(x_{\rm outer})=0\,. 
\end{equation}
We note that these boundary conditions make the differential operator
in equation~(\ref{eq:sl}) self-adjoint: therefore the eigenvalues
$\lambda$ are real, and all slow p-modes are stable.  The complete set
of eigenfunctions also form a complete basis; however, we note that
not all eigenvalues are slow, and thus we do not expect this set to
describe the evolution of arbitrary perturbations, but only the ones
that obey the slow mode condition, $\Omega\gg\vert\omega\vert$.

\begin{figure}
\centering
\includegraphics[width=0.5\textwidth]{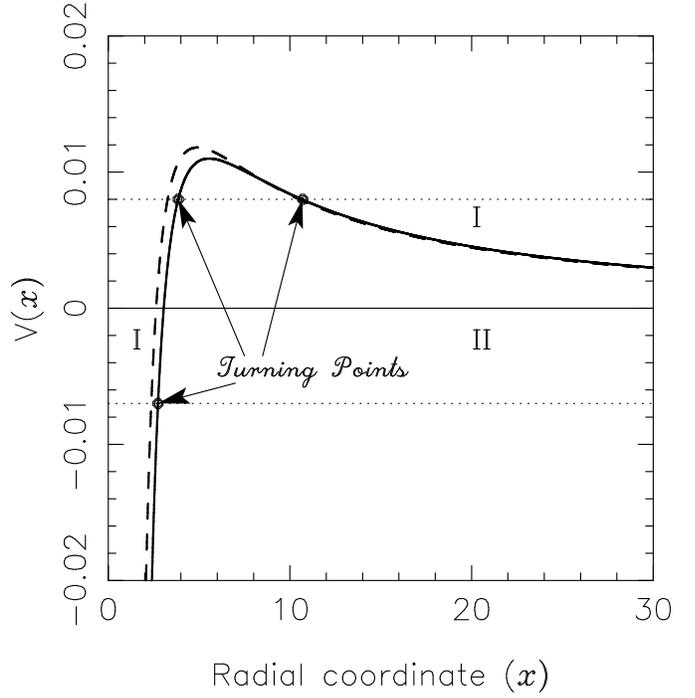}
\caption{The effective potentials for our barotropic approximations to
  the SS disc. The solid line corresponds to ${\rm V}_1(x)$ and the
  dashed line to ${\rm V}_2(x)$, described in \S~4. The positive
  values of $\lambda$ can provide both Type~I and Type~II eigenvalues
  as described in the text. The negative values, although seemingly
  allowing eigenstates lead to no such solution. Note that positive
  values of $\lambda$ correspond to the negative frequency modes.}
\label{fig1}
\end{figure}

\subsection{Effective potential and WKB approximation}

In the usual WKB approximation we substitute the trial solution 
\begin{equation}
\Theta(x) = A(x)\exp\left[\frac{i}{\mu}\int^x \tilde{k}dx\right]\,,
\end{equation}
in the following equation: 
\begin{equation}
\mu^2 \der{}{x}\left({\rm P}(x)\der{\Theta}{x}\right)  +\left({\rm Q}(x) + \lambda {\rm W}(x) \right)\Theta=0\,.
\end{equation}
Here $\mu$ is an ordering parameter which is finally set equal to
unity.  $A(x)$ and $\tilde{k}(x)$ are the amplitude and the wavevector
respectively.  Collecting terms of zeroth order in $\mu$ leads to the
dispersion relation
\begin{equation}
\label{eq:disp} \tilde{k}^2 = \frac{{\rm Q}(x) + \lambda {\rm W}(x)}{{\rm P}(x)}
= \frac{2\Omega}{c_s^2}\left(\pom - \omega\right)\,,
\end{equation}
which is identical to equation~(\ref{eq:drslow}). However, we find
that this dispersion relation, together with the Bohr-Sommerfeld
quantization condition of equation~(\ref {eq:quant}) predicts
eigenvalues that compare poorly with those obtained from numerical
integration of the Sturm-Liouville equation.  Hence we have
reformulated equation~(\ref{eq:sl}), using new variables $\eta(x) =
\sqrt{{\rm P}(x)}$ and $\Psi = \sqrt{{\rm P}(x)}\Theta$. Then
equation~(\ref{eq:sl}) takes the Schr\"odinger-like form
\begin{equation}
\Psi'' + K^2(x)\Psi = 0\,,
\end{equation}
where
\begin{equation}
 K^{2}(x)=\frac{1}{\eta^{2}(x)}\left[{\rm Q}(x) +\lambda{{\rm W}(x)}-\eta(x)\eta''(x)\right]\,,
\end{equation}
which on defining ${\rm V}(x) = ( -{\rm Q}(x) +\eta(x)\eta''(x))/{\rm W}(x)$ can be written as
\begin{equation}
\label{eq:pot} K^{2}(x)=\frac{{\rm W}(x)}{\eta^{2}(x)}\left[\lambda-{\rm V}(x)\right]\,.
\end{equation}
This dispersion relation differs from equation~(\ref{eq:drslow}) and
seems to better describe the numerical solutions, giving a match with
the numerically obtained eigenvalues up to a few per cent, as can be
seen in Table~1.

Note that $K^2(x)$ in equation~(\ref{eq:pot}) differs from the
standard form for the Schr\"odinger equation by the factor ${\rm
  W}(x)/\eta^{2}(x)$. However, it is very useful for discussions of
the turning points, where $K^2(x)=0$, separating classically
accessible regions from the forbidden ones.  The solution is
oscillatory where $K^2>0$, implying $\lambda > {\rm V}(x)$, and
non-oscillatory otherwise.  Since the disc is finite the eigen
spectrum is always discrete and there are two distinct types of
spectra:

\noindent
{\bf Type I}: This occurs when there is at least one turning point
within the disc. In the case of a single turning point we can have
oscillatory solution on either side of the turning point, depending on
the form of $K^2(x)$. If there are more than one turning point then we
could either have oscillatory behaviour confined between the turning
points or outside, such as the case of the SS disc discussed below.

\noindent
{\bf Type II}: This occurs when
there are no turning points within the disc and the discreteness of
the spectra depends entirely on the size of the disc.  

To obtain real eigenvalues we need to consider the possibility of
satisfying $K^2>0$ in a bounded region, which could either be bounded
by one or both of the disc boundaries or by turning points. Thus a
useful first step is to plot this potential for the problem at
hand. If the potential allows regions that can support bound states,
we search for solutions numerically, and to verify our results we use
the WKB approximation.

Let us consider the case of the SS disc shown in Fig~\ref{fig1}.  On
the negative side the potential blows up at the inner edge at $x=1$,
where the perturbations are assumed to vanish. If $\lambda < 0$ is to
be a valid eigenvalue, then it must satisfy the quantization condition
\begin{equation}
\label{eq:WKBQ} \int_1^a K(x)\,dx = \left(n+\frac{3}{4}\right)\pi\,,\qquad n=0,1,2,\ldots
\end{equation}
where, $x=a > 1$ is a turning point which separates a classically
accessible region (I) on the left from a forbidden region (II) on the
right.  The case of positive eigenvalues, $\lambda > 0$, is more
interesting.  If the eigenvalue is positive and smaller than the
maximum value of ${\rm V}(x)$, then there are two turning points (say,
$x=a, b$ with $b > a$) separating three distinct regions.  The
classically forbidden region lies between $x=a$ and $x=b$, separating
the two classically accessible regions $(1, a)$ and $(b, x_{\rm
  outer})$.  If $\lambda$ is greater than the maximum of ${\rm V}(x)$,
then whole disc is classically allowed.  Below we present numerical
results on eigenvalues and eigenfunctions, and use WKB approximation
to understand them. It turns out that WKB approximation is very useful
even for the case of small quantum numbers.
\begin{figure}
\centering
\includegraphics[width=0.5\textwidth]{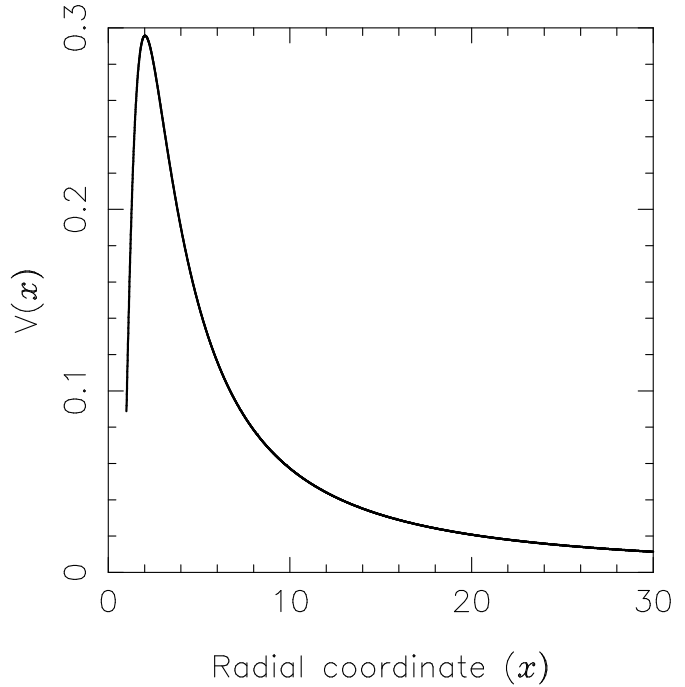}
\caption{The effective potential for the Kuzmin disc. Positive 
values of $\lambda$ lead to a discrete spectrum of both Type~I and
Type~II eigenvalues. There are no negative eigenvalues.}
\label{fig2}
\end{figure}

\section{Numerical results}
As discussed in the last section, we consider finite disc and we
expect the spectra of equation~(\ref{eq:sl}) to be discrete. These
modes would have observational consequences since they would rotate at
a definite frequency around the disc. The perturbations in the
enthalpy would lead to azimuthal variations in the temperature and
density across the disc, which might be observable depending on the
amplitude of perturbations.

Since very little is known about the surface density profiles of the
discs we carry out a simplistic calculation based on certain standard
forms of the disc as test cases.  We also consider the generic power
law profile. Some of the profiles considered below are formally
infinite in size, however as noted above, we need to keep the disc
finite. This would imply that the surface density would abruptly fall
to zero at the outer disc radius. This is unphysical and we expect
that there would be a thin transition region that would deviate from
the density profile being considered near $x_{\rm outer}$. If the
eigenvalues and eigenmodes are not very sensitive to $x_{\rm outer}$,
then this is not an issue and in our numerical investigation we indeed
find this to be true. In all the examples below $x_{\rm outer} = 50$.
In Table~1 we give the first a few eigenvalues for $x_{\rm outer} = 50$ and 
$x_{\rm outer} = 200$, as $\lambda_{50}$ and $\lambda_{200}$ respectively.
It can be noticed that eigenvalues do not change substantially.

\begin{figure}
\centering
\includegraphics[width=0.5\textwidth]{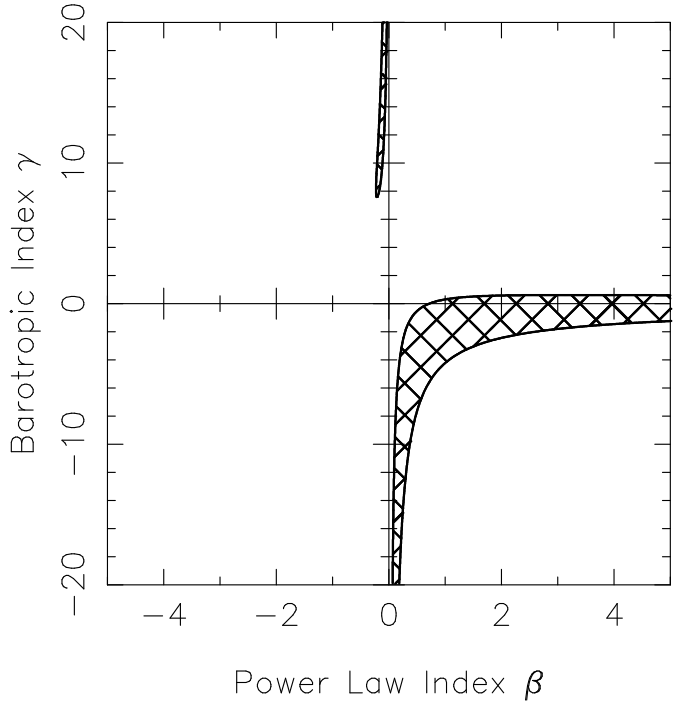}
\caption{The allowed region (hatched) in the $\beta$-$\gamma$ space to obtain
Type~I negative eigenvalues (corresponding to the positive frequency
modes) for the power law discs (case (ii)). Most of the region is either unphysical or uninteresting.}
\label{fig3}
\end{figure}

{\bf Shakura-Sunyaev (SS) discs}: We first consider the standard model
of an accretion disc proposed by Shakura and Sunyaev \citep{ss73}. The
surface density and temperature of this disc are given by (see
\citet{fkr02}):
\begin{eqnarray}
\Sigma_{\rm SS} &=& 5.2\,\alpha^{-4/5}{\dot{M}}_{16}^{7/10}{m_1}^{1/4}{R_{10}}^{-3/4}f^{14/5}
{\text{~g~cm}}^{-2}\,,
\label{eq:sigmass}\\[1em]
T_{\rm SS} &=& 1.4\times 10^4\alpha^{-1/5}{\dot{M}}_{16}^{3/10}{m_1}^{1/4}{R_{10}}^{-3/4}f^{6/5}\,{\rm K}\,,
\label{eq:tempss}
\end{eqnarray}
where 
\begin{equation}
f =\left[1-\left(\frac{R_*}{R_{10}}\right)^{1/2}\right]^{1/4}\,.
\end{equation}
${\dot{M}}_{16}$ is the mass accretion rate in the units of $10^{16}$g
s$^{-1}$, $m_1$ is the mass of disc in solar mass units, $R_{10}$ is
radius in the units of $10^{10}$cm, $R_*$ is the radius of the central
object in the units of $10^{10}$cm, and $\alpha$ is the Shakura-Sunyaev viscosity parameter.

Although the SS disc is not based on a barotropic model, we find that
a barotropic disc with index $\gamma=2$ serves as a reasonable
approximation.  We have considered two cases:

\begin{enumerate}
\item Choosing $\Sigma_0 = \Sigma_{\rm SS}$ of
equation~(\ref{eq:sigmass}), and deriving the temperature profile for
$\gamma=2$, we find that the effective potential is
\begin{equation}
{\rm V_1}(x)= \frac{344 - 590\sqrt{x}+225x}{800\,x^{9/4}\,(1-\sqrt{1/x})^{13/10}}\,.
\end{equation} 

\item Choosing $T_0 = T_{\rm SS}$ of equation~(\ref{eq:tempss}), and
deriving the surface density profile for $\gamma=2$, we find that the
effective potential is
\begin{equation}
{\rm V_2}(x)= \frac{12 - 22\sqrt{x}+ 9x}{32\,x^{9/4}\,(1-\sqrt{1/x})^{3/2}}\,,
\end{equation} 
\end{enumerate}
\noindent
where we have used the natural length scale given by the inner disc
radius, which we have adopted for conversion of our eigen equation
into a dimensionless from. As Fig~(\ref{fig1}) shows, ${\rm{V}_1}(x)$
and ${\rm{V}}_2(x)$ are quite similar to each other.  The effective
potential blows up at the inner edge of the disc and steadily climbs
up above zero and then decreases asymptotically.  In the immediate
vicinity of the inner edge a negative $\lambda$ gives an oscillatory
solution, and to the right of it is a classical turning point. Beyond
the turning point the solution is exponentially decaying. This
suggests that discrete, Type~I, negative eigenvalues might exist,
however, both the numerical search and the WKB quantitation condition
(\ref{eq:WKBQ}), fail to find such discrete eigenvalues. For
$0<\lambda\lesssim 0.01$, we find discrete, Type~I eigenvalues for
which the oscillatory behaviour is outside the region bounded by two
turning points. For $\lambda\gtrsim 0.01$, we find discrete
eigenvalues of Type~II, where the separation between neighbouring
eigenvalues decreases with increasing values of the outer radius of
the disc.

{\bf The Kuzmin Disc}: In contrast to the SS discs, the Kuzmin disc
has a centrally concentrated surface density, and hence offers a
distinct case in which to study slow modes. The surface density
profile in this case is given by
\begin{equation}
\Sigma(R) = \frac{aM_D}{2\pi\,(R^2+a^2)^{3/2}}\,,
\end{equation}
where $M_D$ is the mass in the disc and $a$ is the core radius. The
surface density extends all the way to $R=0$. If we take $a$ as the
size of the inner radius of the disc then we can rescale our equation
by choosing $R_{\star}=a$, leading to the effective potential
\begin{equation}
{\rm V}(x)=\frac{3(x^4(5-4\gamma+3\gamma^2)+x^2(6-8\gamma)+1)}{8x^{1/2}(1+x^2)^{(3\gamma+1)/2}}.
\label{eq:potkuz}
\end{equation}
In Fig~(\ref{fig2}) we have plotted this potential for $\gamma=4/3$.
As may be seen, this case qualitatively resembles Fig~(\ref{fig1}),
and the discussion in \S~2.4 carries through. For negative values of
$\lambda$, we can infer from Fig~(\ref{fig2}) that $K^2(x)<0$, so wave
like solutions are not possible. For positive values of $\lambda$ in
the range $0<\lambda\lesssim 0.3$, we can have Type~I eigenvalues, but
this region of $\lambda$ is further divided into two parts: for $0<
\lambda \le {\rm V}(1)$, there is only one turning point and
oscillatory behaviour is possible for radius greater than the turning
point, and for ${\rm V}(1)<\lambda\lesssim 0.3$, there are two turning
points and oscillatory behaviour is possible outside the region
bounded by the two turning points. For $\lambda\gtrsim 0.3$ the
eigenvalues are of Type~II. This behaviour is confirmed by numerical
integration of the eigenvalue equation.  We can also admit values of
$\gamma$ other than $4/3$. However, the effective potential in
equation~(\ref{eq:potkuz}) retains the general shape of
Fig~(\ref{fig2}), and the conclusions stated above remain valid.

\begin{figure}
\centering
\includegraphics[width=0.5\textwidth]{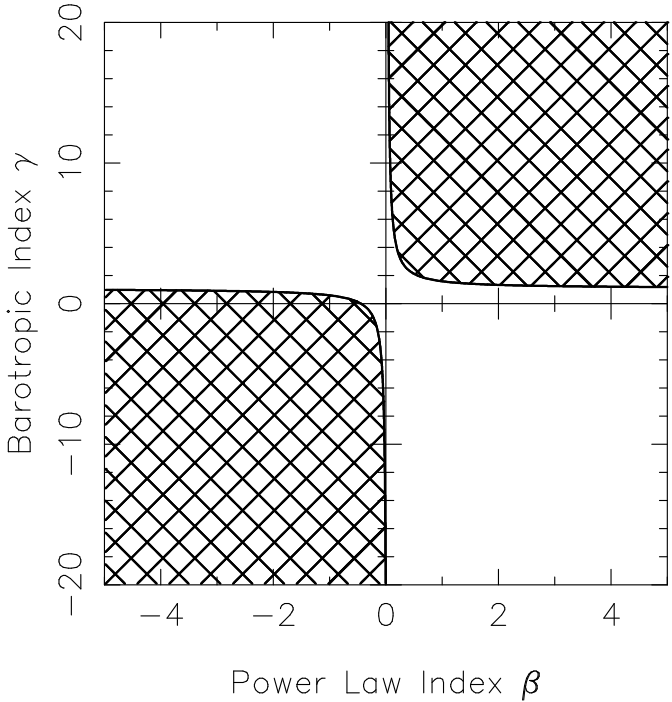}
\caption{The allowed region (cross-hatched) in the $\beta$-$\gamma$ space to obtain
discrete positive eigenvalues (corresponding to the negative frequency
modes), corresponding to Type~I eigenvalues for case (iii), for the power law discs.}
\label{fig4}
\end{figure}

\begin{figure}
\centering
\includegraphics[width=0.5\textwidth]{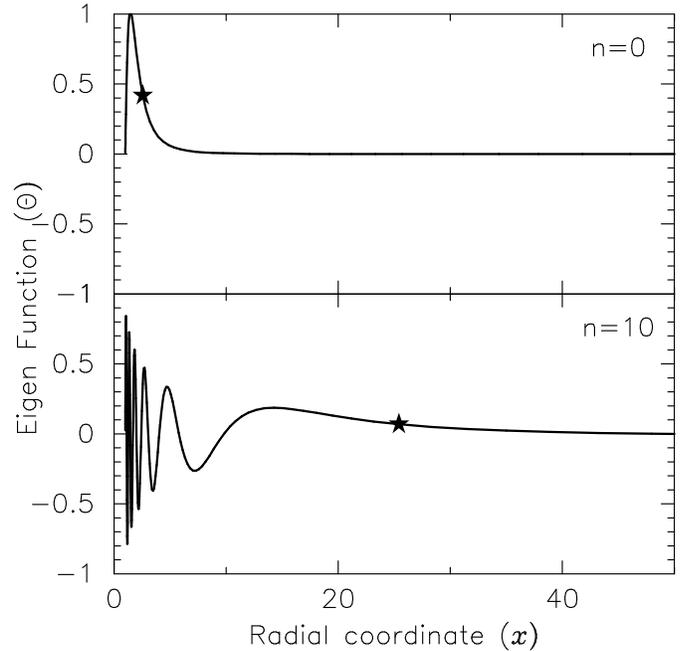}
\caption{Two eigenfunctions (Type-I eigenvalue) obtained for the power law case (case iii) with
$\beta=\gamma=2$. The higher quantum number leads to a more radially
extended eigenmode. Star marks the position of the turning point.}
\label{fig5}
\end{figure}

\begin{figure}
\centering
\includegraphics[width=0.5\textwidth]{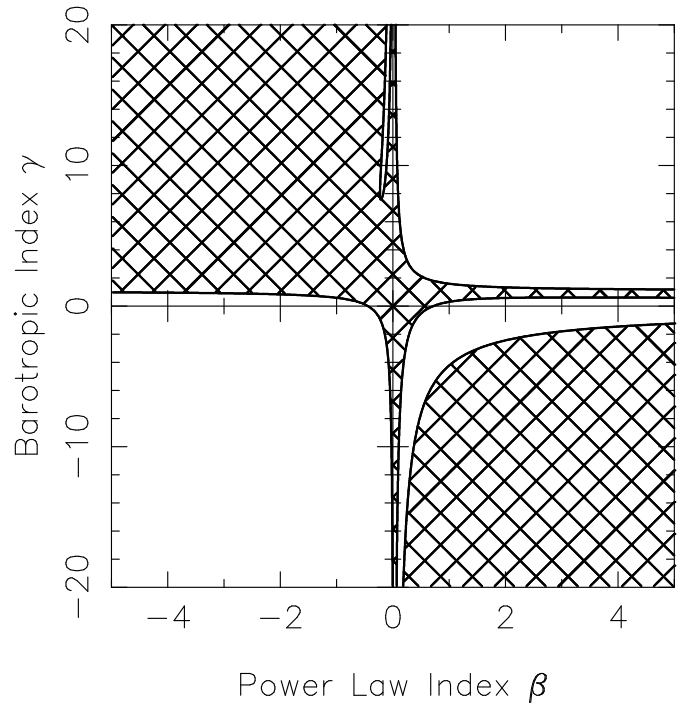}
\caption{The allowed region (hatched) in the $\beta$-$\gamma$ space for the
positive Type~I eigenvalues (corresponding to negative frequency eigenmodes), for case (iv)
for the power law discs.}
\label{fig6}
\end{figure}

\begin{figure}
\centering
\includegraphics[width=0.5\textwidth]{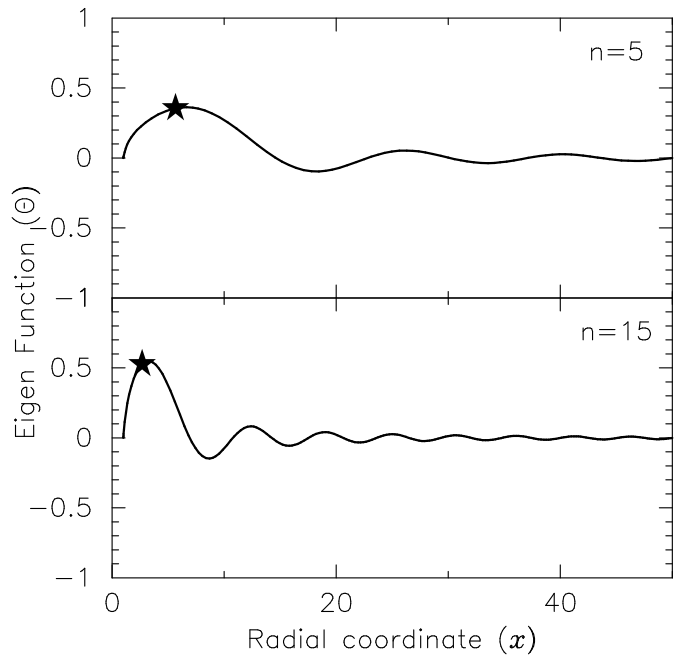}
\caption{Two eigenfunctions (Type-I eigenvalue) obtained for the power law case (case iv)
with $\beta= -2, \gamma=2$. Star marks the position of the turning
point.}
\label{fig7}
\end{figure}

{\bf Power Law Discs}: Certain physical models (e.g. \citet{ny94})
naturally lead to scale invariant discs that follow a power law
profile. Although in these models there is no associated length scale,
we choose to truncate the disc at the inner disc radius thus leading
to the form
\begin{equation}
\Sigma_0=\Sigma_{\star}\,x^{\beta}\,,
\end{equation} 
where, $\Sigma_{\star} = \Sigma(R_{\star})$ and $\beta$ is the power law
index. The potential is given by
\begin{equation}
{\rm V}(x)= C(\beta,\gamma) x^{\nu}\,,
\end{equation}
where
\begin{eqnarray}
C(\beta,\gamma) &=& \frac{1}{8}\left(3 + \beta\left(-4 + \gamma(4 + \gamma\beta)\right)\right)\,,\\
\nu(\beta,\gamma)&=& \gamma\beta - \beta-1/2\,. 
\end{eqnarray}

There are four distinct possibilities:
\begin{enumerate}
\item$C(\beta , \gamma)< 0$, $\nu(\beta , \gamma)<0$: The region in
  $\beta$-$\gamma$ space satisfying these conditions is plotted in 
  Fig~(\ref{fig3}). The regions where these conditions are satisfied
  are those with (1) $\beta$ positive and $\gamma$ negative or (2)
  very small negative $\beta$ and very large $\gamma$. Both these
  cases are unphysical.  Note that there is a region with
  large, positive $\beta$, and small, positive $\gamma$, and these are
  also physically uninteresting.
\item$C(\beta , \gamma)< 0$, $\nu(\beta , \gamma)>0$: It can be verified that for 
$\nu(\beta , \gamma)>0$ $C(\beta , \gamma)$ is always positive and
hence it is impossible to satisfy these conditions.
\item$C(\beta , \gamma)> 0$, $\nu(\beta , \gamma)>0$: These two
  constraints give us a region in the $\beta$-$\gamma$ space,
  displayed in Fig~(\ref{fig4}), that admits physically reasonable
  values. For ${\rm V}(x_{\rm inner})<\lambda< {\rm V}(x_{\rm
  outer})$, we have one turning point admitting Type~I eigenvalues.
  In Fig~(\ref{fig5}) we plot two examples of eigenmodes for the case
  $\beta=\gamma=2$. The eigenfunctions for small quantum numbers are
  found to be centrally concentrated while they extend to larger radii
  for larger quantum numbers. It should be noted that our solutions
  are regular at the turning points\footnote{The positions of the ILRs,
  obtained from,
  $ \lambda = -{\rm{Q}(x)}/{\rm{W}(x)}$ are slightly different from 
  the turning points obtained from the equation
 $ \lambda = -(\rm{Q}(x) - \eta(x)\eta''(x))/{\rm{W}(x)}$.}.
  The first few eigenvalues for this
  case are tabulated in Table~1, where we find an excellent match with
  the WKB eigenvalues.  Outside this range of positive $\lambda$ there
  are no turning points and hence only Type~II eigenvalues are
  possible.
\item$C(\beta , \gamma)> 0$, $\nu(\beta , \gamma)<0$: The region in 
 $\beta$-$\gamma$ space satisfying these constraints is plotted in Fig~
 (\ref{fig6}). For $0<\lambda< {\rm V}(1)$, we have one turning point
 admitting Type~I eigenvalues. Here the classically accessible region
 is bounded by the turning point on left and $x_{outer}$ on
 right. Numerical solution for $\beta = -2$ and $\gamma = 2$ are
 plotted in Fig~(\ref{fig7}). Comparison with Fig~(\ref{fig5}) shows
 that the solutions are more radially extended, with well separated
 peaks. For $\lambda > {\rm V}(1)$ there is no turning point and the
 eigenvalues are of Type~II.
\end{enumerate}

\section{Discussion and Conclusions}

We have presented a theory of slow $m=1$ linear pressure modes (or
``p-modes'') in thin accretion discs around massive compact objects,
such as white dwarfs and neutron star. These modes are enabled by the
small deviation from a purely Keplerian flow, due to fluid pressure
rather than disc self-gravity. For simplicity we have taken the fluid
to be barotropic. Our formulation largely follows that of
\citet{tre01}, although there is a key difference: using the WKB
approximation, \citet{tre01} argued that fluid discs for which disc
self-gravity dominates fluid pressure can support slow modes, if the
Mach number ${\mathscr M}$ is much larger than the Toomre $Q$ parameter. This
condition may be satisfied in relatively cool discs, but not for thin
accretion discs around white dwarfs or neutron stars. In these discs,
$Q\gg {\mathscr M}\gg 1$, and the analysis in \citet{tre01} does not
apply, because disc self-gravity is negligible when compared with
fluid pressure in thin accretion discs.  This implies that the
precession rate of the apsides $(\pom)$ of a fluid element in a nearly
circular orbit is determined by the fluid pressure; to order of
magnitude, $\pom \sim (\Omega/{\mathscr M}^2)$.

\begin{table}
\centering
\label{tab:comp}
\begin{center}
\begin{tabular}{cccc} \hline
$n$ & WKB $\lambda$  & Numerical $\lambda_{50}$ & Numerical $\lambda_{200}$ \\\hline
0 & 13.42  &  13.86  &  13.86\\
1 & 30.15  &  30.73  &  30.73\\
2 & 52.31  &  53.06  &  53.06\\
3 & 79.99  &  80.92  &  80.91\\
4 & 113.22 & 114.32  &  114.31\\
5 & 151.99 & 153.29  &  153.26\\
6 & 196.31 & 197.83  &  197.75\\
7 & 246.18 & 247.96  &  247.80\\
8 & 301.60 & 303.72  &  303.40\\
9 & 362.57 & 365.16  &  364.54\\
10&  429.09& 432.31  &  431.24\\\hline
\end{tabular}
\caption{The eigenvalues for the power law model with \mbox{$\beta=\gamma=2$}
  are tabulated here for comparison between those obtained numerically
  and those obtained using the WKB approximation. The columns
  $\lambda_{50}$ and $\lambda_{200}$ are the eigenvalues 
  corresponding to $x_{\rm{outer}} = 50$ and $200$ respectively. The match between
  numerical and WKB eigenvalues is within
  a few per cent, and remains good even for small quantum numbers. The eigenvalues
  are not very sensitive to precise value of $x_{\rm{outer}}$}.
\end{center}
\end{table}

A WKB analysis was used first to argue that thin accretion discs can
support large-scale $(k(R) R \sim 1)$, $m=1$ p-modes with small
angular frequencies, $\omega\sim\pom \sim (\Omega/{\mathscr M}^2)$. As
noted by \citet{tre01}, these long wavelength modes may dominate the
appearance of the disc, and are not expected to be damped by
viscosity.  We derived an eigen equation for slow linear modes and
showed that it is identical to a Sturm-Liouville problem. The
differential operator being self-adjoint implies that the eigenvalues
are all real, so that all slow p-modes are stable. This corresponds to
the result in \citet{tre01} that all slow modes of the softened
gravity disc are stable.  We solved the Sturm-Liouville problem
numerically for a variety of unperturbed discs, and summarise our
results below.

\begin{enumerate}
\item The first corresponds to two different kinds of barotropic
approximations to the Shakura-Sunyaev thin disc, which have modes with
negative eigenfrequencies.

\item The second is the Kuzmin disc, which is more centrally
concentrated. As earlier, this too supports only
negative eigenfrequencies.

\item Power-law discs can support modes with negative eigenfrequencies for reasonable 
values of $\beta$ and $\gamma$. For certain combinations of these
parameters power law discs can support positive eigenfrequencies as well;
however the range of parameters turn out to be physically
uninteresting.

\end{enumerate}

If slow modes are stable, it is necessary to consider how they could
be excited. Since they have azimuthal wavenumber $m=1$, we need look
for excitation mechanisms which possess the same symmetry, at least in
the linear limit. One possibility is from the stream of matter from
the secondary star that feeds the accretion disc. When viewed in the
rest frame of the primary, the region where the stream meets the outer
edge of the disc rotates in a prograde sense with angular frequency
equal to the orbital frequency of the binary system.  Since the
orbital frequency of the binary can be much smaller than the orbital
frequency of the gas in the accretion disc, there is the possibility
of the resonant excitation of a slow mode if it has positive
frequency. In the cases we have considered, we find only negative frequencies
belonging to a discrete spectrum, allowing only for non--resonant driving. 

There is, however, another alternative that does not rely on external
sources of excitation.  \citet{zl05} have studied linear waves in thin
accretion discs and applied their theory to slow $m=1$ modes around
black holes. They used the pseudo-Newtonian gravitational potential of
\citet{pw80} to model the general relativistic effects due to a
Schwarzschild black hole. In this case $\pom$ is due to the deviation
of the pseudo-Newtonian potential from a Kepler potential, so their
slow modes are not driven by pressure (as is true in all the cases we
have considered); hence more detailed comparison with our work is not
possible. What is interesting is that they find that their slow modes
to have negative energy and angular momentum, and suggest that slow
modes may be excited spontaneously through the action of viscous
forces.  This possibility should be examined in the context of the
p-modes we have studied. However, this would require reformulating the
eigenvalue problem taking into account viscous effects.

\def\etal{{\it et~al.}}
\def\apj{{Astroph.\@ J. }}
\def\mnras{{Mon.\@ Not.\@ Roy.\@ Ast.\@ Soc.}}
\def\aap{{Astron.\@ Astrophys.}}
\def\aj{{Astron.\@ J.}}
\def\apss{{Astroph.\@ Space \@ Science}}
\def\pasj{{Publ.\@ Astron.\@ Soc.\@ Japan}}
 
\end{document}